\documentstyle[aps,prl]{revtex}
\begin{document}
\input epsf.sty

\twocolumn[\hsize\textwidth\columnwidth\hsize\csname %
@twocolumnfalse\endcsname

\title{Spin Dependence of Correlations in Two-Dimensional
Quantum Heisenberg Antiferromagnets}

\author{N. Elstner$^1$, A. Sokol$^2$, R.R.P. Singh$^3$,
M. Greven$^4$, and R.J. Birgeneau$^4$}
\address{$^1$Service de Physique Th\'{e}orique, CEA-Saclay, 91191
Gif-sur-Yvette Cedex, France}
\address{$^2$Department of Physics, University of Illinois at
Urbana-Champaign, Urbana, IL 61801-3080\\
and L.D. Landau Institute, Moscow, Russia}
\address{$^3$Department of Physics, University of California, Davis, CA
95616}
\address{$^4$Department of Physics, Massachusetts Institute of Technology,
Cambridge, MA 02139}

\date{February 27, 1995}

\maketitle

\begin{abstract}
We present a series expansion study of spin-S square-lattice Heisenberg
antiferromagnets. The numerical data are in excellent agreement with recent
neutron scattering measurements. Our key result is that the correlation length
$\xi$ for $S>1/2$ strongly deviates from the exact $T\to 0$
(renormalized classical, or RC) scaling prediction for all
experimentally and numerically accessible temperatures. We note basic trends
with S of the experimental and series expansion correlation length data and
propose a scaling crossover scenario to explain them.
\end{abstract}

\phantom{.}
]

\narrowtext

In recent years much attention has focused on
two-dimensional (2D) square-lattice quantum Heisenberg
antiferromagnets, described by the Hamiltonian
\begin{equation}
H = J \sum_{\langle ij\rangle} {\bf S}_i {\bf S}_j,
\label{H}
\end{equation}
where $\langle ij\rangle$ denotes summation over all pairs of nearest
neighbors.
In their seminal work \cite{CHN}, Chakravarty, Halperin, and Nelson
(CHN), utilizing a mapping of the low-energy spectrum of Eq.(\ref{H})
onto the quantum non-linear sigma model (QNL$\sigma$M),
have shown that the low-temperature properties of these systems obey
renormalized classical (RC) scaling, where the correlation length
$\xi\approx 0.37\times
(c/\rho_s) \exp(2\pi\rho_s/T)$ for $T\ll \rho_s$ (here
$\rho_s$ is the $T\to0$ spin stiffness, and $c$ is the spin-wave velocity).
Subsequently,
Hasenfratz and Niedermayer calculated for the 2D QNL$\sigma$M
the exact value of the prefactor and
the leading $O(T/2\pi\rho_s)$ correction \cite{HN}:
\begin{equation}
	\xi_{\rm HN} = \frac{e}{8} \frac{c}{2\pi \rho_s}
	 \exp\left(\frac{2\pi \rho_s}{T}\right)\, \left[ 1 -
	 \frac{T}{4\pi\rho_s} + O\left(\frac{T}{2\pi\rho_s}\right)^2\right],
	\label{HN}
\end{equation}

The neutron scattering measurements of $\xi(T)$
in $S=1/2$ layered Heisenberg
antiferromagnets such as La$_2$CuO$_4$ \cite{Keimer:lco} and
Sr$_2$CuO$_2$Cl$_2$ \cite{Greven:scoc} reveal a remarkable agreement
with Eq.(\ref{HN}).
At first sight, one would expect the RC description to improve
as the value of the spin increases. If $S$ is formally regarded as
a continuous variable, then the N\'eel order is expected to vanish for
some $S<1/2$. At the critical point, where $\rho_s$ vanishes,
Eq.(\ref{HN}) fails and the correlation length is, in fact,
inversely proportional to the temperature; this is the quantum critical
regime (QC), discussed in Ref. \cite{CHN,CSY}
(it was argued earlier \cite{CSY,SGS} that the $S=1/2$ Heisenberg model
(\ref{H}) exhibits certain signatures of QC behavior for $T>0.5J$;
we do not discuss
RC to QC crossover effects in this letter).
Naively,
increasing the value of spin moves the system away from this limit so
that the RC behavior would be more pronounced.

\begin{figure}
\centerline{\epsfxsize=3.0in\epsfbox[35 90 585 825]{s1.ps}}
\caption{
Semi-log plot of the series expansion result for the correlation length in the
$S=1$
Heisenberg antiferromagnet versus $T/J$ (solid line)
together with the
experimental data for K$_2$NiF$_4$ (solid circles,
\protect\cite{Greven:scoc}) and
La$_2$NiO$_4$ (open circles, \protect\cite{Greven:S=1}).
Numerical and experimental results agree with each other,
and all deviate from the exact RC prediction \protect\cite{HN}
evaluated using known values for $c$ and $\rho_s$
\protect\cite{1/S,Singh:rhos} (dashed line).
}
\label{fig:s1}
\end{figure}

However, such an expectation does not
hold. Recently,
Greven and co-workers \cite{Greven:scoc,Greven:S=1}
have reported a significant discrepancy between the
neutron scattering measurements of the correlation length in the
$S=1$ systems K$_2$NiF$_4$ and La$_2$NiO$_4$ and the RC prediction.
Preliminary experiments on the $S=5/2$ system Rb$_{2}$MnF$_{4}$ reveal an even
larger discrepancy \cite{Greven:S=5/2}.
As is evident from Fig.\ref{fig:s1}, our series expansion result for $S=1$
\cite{EGSS}
is in excellent agreement with the experimental data in the region of
overlap, and also deviates substantially from the RC prediction
(note, that $\xi$ is expressed in units of the lattice constant $a$).

In order to investigate the origin of this deviation, we have calculated
high-temperature expansions for the Fourier transform of the
spin-spin correlation function
$\langle S^z_{-{\bf q}} S^z_{\bf q}\rangle$
for all spins in the range $S=1/2$ to $S=5/2$.
The generation of the series expansions will be discussed elsewhere.
We present the
data for the $T=0$ structure factor $S_0 = S({\bf Q})$, where
${\bf Q}=(\pi/a,\pi/a)$ is the staggered
ordering wavevector,
and the second moment correlation length $\xi$ is defined by
$\xi^2 =  - \left.\left(\partial^2 S({\bf Q}+{\bf
q})/\partial q^2\right)\right|_{q=0}/2S_0$.
The series are analyzed by either performing direct Pad\'e approximation
or by taking the logarithm of the
series first and then calculating the Pad\'e approximants. The latter is
likely to show better convergence if the correlation length increases
exponentially fast at low temperatures. We restrict
ourselves to temperature ranges where different methods of extrapolation
agree within a few per cent.

\begin{figure}
\centerline{\epsfxsize=3.0in\epsfbox[50 90 600 820]{ratio.ps}}
\caption{
Series expansion results for the correlation length plotted as
$\xi/\xi_{\rm HN}$ versus $T/\rho_s$. Here $\xi_{\rm HN}$ is the
exact RC prediction \protect\cite{HN}, Eq.(\protect\ref{HN}),
with $c$ and $\rho_s$ from Ref.\protect\cite{1/S,Singh:rhos}.
The ratio $\xi/\xi_{\rm HN}$, which is fairly
close to unity for $S=1/2$, gradually decreases as $S$ increases,
suggesting increasing disagreement with RC theory for larger $S$.
}
\label{fig:ratio}
\end{figure}

The ratio between the calculated
$\xi$ and the Hasenfratz and Niedermayer formula
$\xi_{\rm HN}$ given by Eq.(\ref{HN}), is
plotted in Fig.\ref{fig:ratio} for different spins
as a function of $T/\rho_s$. Contrary to the naive expectation that
the RC behavior becomes more pronounced as $S$ increases, we find that
for the range of temperatures probed here $\xi$
monotonically  deviates from $\xi_{\rm HN}$ as $S$
increases. As noted above, this same systematic trend is seen
experimentally \cite{Keimer:lco,Greven:scoc,Greven:S=1,Greven:S=5/2}.

\begin{figure}
\centerline{\epsfxsize=3.0in\epsfbox[30 60 585 795]{Sxi.ps}}
\caption{
Semi-log plot of $S\xi/a$ versus $T/JS^2$ of
Sr$_2$CuO$_2$Cl$_2$ ($S=1/2$, open circles, \protect\cite{Greven:scoc})
and K$_2$NiF$_4$ ($S=1$, solid squares, \protect\cite{Greven:scoc})
together with the series expansion results
($1/2\leq S\leq 5/2$, represented by the various lines) as well as
Monte Carlo data ($S=1/2$, solid squares, \protect\cite{Greven:mc}).
However impressive, this data collapse is inconsistent with
Eq.(\protect\ref{HN}) with
$c$ and $\rho_s$ from Ref.\protect\cite{1/S}.}
\label{fig:Sxi}
\end{figure}

In an attempt to understand these data,
we have considered various possible scaling scenarios. To highlight
the expected spin dependence, we note that for a square-lattice
nearest-neighbor Heisenberg antiferromagnet,
$\rho_s = Z_\rho(S) JS^2$ and $c = 2^{3/2} a Z_c(S) JS$. The quantum
renormalization factors $Z_\rho(S)$ and $Z_c(S)$ are known from
spin-wave theory \cite{1/S} (for all values of $S$), $T=0$ series
expansion \cite{Singh:rhos} (for $S=1/2$ and $S=1$), and
Monte Carlo studies \cite{Wiese:Ying} (for $S=1/2$).
For $S=1/2$ and $S=1$ these different methods yield good agreement.

Eq.(\ref{HN}) may then be written as
\begin{equation}
\frac{S\xi}{a}  =
\frac{e Z_c}{2^{5/2}\pi Z_\rho}
\exp\left(\frac{2\pi Z_\rho}{T/JS^2}\right)
\left[1 - \frac{T/JS^2}{4\pi Z_\rho} + \ldots \right],
\label{Sxi}
\end{equation}
which suggests
plotting $S\xi/a$ versus $T/JS^2$ to elucidate the
dependence on $S$.
The results so-obtained are shown in Fig.\ref{fig:Sxi}.
For $S=1/2$ we also show Monte Carlo data \cite{Greven:mc},
which extend to somewhat lower
temperatures than the series expansion result.
Surprisingly, over the range of $S\xi/a$ from 1 to at least 50,
the data to a good approximation fall on the same curve.
For the sake of clarity, we have omitted experimental data for
La$_{2}$CuO$_{4}$ ($S=1/2$), La$_{2}$NiO$_{4}$ ($S=1$), and Rb$_{2}$MnF$_{4}$
($S=5/2$), all of which fall on the approximate ``scaling''
curve of Fig. \ref{fig:Sxi} to within experimental error.

It was demonstrated in Ref.\cite{Greven:scoc}
that Eq.(\ref{Sxi}) describes the correlation
length data in absolute units in the $S=1/2$ system Sr$_{2}$CuO$_{2}$Cl$_{2}$
extremely well. Interpreted naively, Fig.\ref{fig:Sxi}
would then suggest that RC behavior holds for all
S, but with quantum renormalization factors $Z_\rho(S)$ and
$Z_c(S)$ that are nearly S-independent for $1/2\leq S\leq 5/2$ and
close to their values at $S=1/2$, $Z_\rho(1/2)\approx0.72$ and
$Z_c(1/2)\approx 1.2$.
{\em However, this is very unlikely.}
Spin-wave theory \cite{1/S} predicts a substantial S-dependence:
$Z_\rho(5/2)\approx 0.95$ and $Z_c(5/2)\approx 1.03$, which
is already close to the classical limit ($S\to\infty$) in which
$Z_\rho$, $Z_c\to1$.
Serious errors in
these values seem very unlikely, given their good agreement for
$S=1/2$ and $S=1$ with those obtained by other methods
\cite{Singh:rhos,Wiese:Ying}.

In an attempt to understand these data,
it is helpful to recall that at any
fixed $T/\rho_s$ RC theory will inevitably
fail for sufficiently large values
of spin. Indeed, a straightforward application of Eq.(\ref{HN}) to the
$S\to\infty$ limit taken at $T/JS^2\sim 1$ would
predict $\xi=0$.
However, the above limit corresponds to the classical
Heisenberg magnet, where $\xi/a$ is known to be
finite and of order unity for $T\sim JS^2$.

One can understand
{\em where} Eq.(\ref{HN}) may fail by following CHN in their derivation of
the leading asymptotic behavior in the RC regime, but taking into
account that $S$ may be large.
CHN have shown that for $T\ll
\rho_s$ the magnetic correlations can be calculated using
classical dynamics, except that all wavevector integrations should be
limited to $|{\bf q}|\alt q_c = T/c$
rather than taken over the whole Brillouin
zone. Here the words ``classical dynamics'' simply mean that for
$|{\bf q}|\alt q_c$, all Bose factors for spin-waves can be
approximated assuming $cq\ll T$.
The key result of CHN is
to show that such a calculation yields the correct spin correlations
for the quantum Heisenberg model.
We now evaluate the cutoff wavevector as
\begin{equation}
q_c \sim \frac{T}{c} \sim \frac{\rho_s}{c} \frac{T}{\rho_s} \sim
\frac{S}{a} \frac{T}{\rho_s}.
\end{equation}
This dependence of $q_c$ on $S$ arises because
the spin stiffness is proportional to the square of $S$, but the
spin-wave velocity is only linear in $S$.

For $S\gg 1$ and $T\sim\rho_s\sim JS^2$,
the cutoff wavevector $q_c\sim S/a\gg\pi/a$ is {\em
outside} the Brillouin zone. Hence, the requirement that $cq\alt T$, or
equivalently $q\alt q_c$,
places no further restrictions on the q-integrations which are already
limited by the Brillouin zone. In this case all of the integrals are
the same as those of the classical Heisenberg magnet, and the
classical $S\to\infty$ limit is recovered.

The crossover temperature $T_{\rm cr}$ between the RC and classical regimes
depends on $S$, and its order of magnitude can be estimated as the
temperature where $q_c\sim a^{-1}$. This yields
\begin{equation}
T_{\rm cr} \sim \frac{c}{a} \sim JS, \ \ \mbox{ while $\rho_s\sim JS^2$}.
\label{Tcross}
\end{equation}
By substituting $T_{cr}$ into Eq.(\ref{HN}),
one concludes that the crossover from RC
behavior at low temperatures to classical
behavior at higher temperatures should occur for
a $\xi_{\rm cr} = \xi(T_{cr})$ that is larger for larger $S$.

\begin{figure}
\centerline{\epsfxsize=3.0in\epsfbox[30 93 585 820]{xi.ps}}
\caption{
Semi-log plot of the correlation length obtained from series expansion plotted
versus $T/(JS(S+1))$ for $S=1/2,1,$ and $5/2$.
For larger spins $\xi/a$ is close to the classical ($S\to\infty$) limit,
which provides evidence that classical ($S\to\infty$) magnetic
behavior holds for  $JS\ll T\ll JS^2$, in agreement with our proposed
scaling crossover scenario.
Note, that in most of the temperature range shown,
the $S\to\infty$ model is
not in the scaling limit and its correlation length deviates
from the expected
$T\to 0$ behavior
$\xi/a\approx \mbox{const}\times (T/JS^2)\exp(2\pi JS^2/T)$.}
\label{fig:xi}
\end{figure}

In order to test this scenario, we plot the correlation length
as a function of $T/(JS(S+1))$ in Fig.\ref{fig:xi},
where $JS(S+1)$ is the {\em classical}
(not $T=0$) spin stiffness. In replacing $S^2$ by
$S(S+1)$, we follow a purely empirical observation that the
correlation length at $T\gg JS$
for $S>3/2$ seems to depend on $S$ primarily
through the combination $S(S+1)$ (for $T\ll JS$,
$\xi$ depends on $S$ only through $\rho_s$ and $c$).
We find that $\xi$ gradually
approaches the classical result as $S$ increases, and the difference
between $\xi$ for the largest spin in our study ($S=5/2$) and
the classical antiferromagnet ($S\to\infty$) is already very small.
This result supports the hypothesis that the deviations from
asymptotic RC behavior evident in Figs.\ref{fig:s1} and \ref{fig:ratio} are
driven primarily by RC to classical crossover effects.

Another important observable of the scaling theory is the Lorentzian
amplitude of the spin-spin correlator, $S_0/\xi^2$,
where $S_0$ is the correlator magnitude at
$Q=(\pi/a,\pi/a)$. This ratio has only a power-law temperature
dependence, and is therefore less sensitive to the model parameters
than $\xi$ or $S_0$ separately. A detailed discussion of this
quantity is beyond the scope of this publication. However, we note some
important issues which need to be clarified before the scaling
behavior of $S_0/\xi^2$ can be fully understood.

The RC scaling prediction for this quantity is \cite{CHN}
\begin{equation}
\frac{S_0}{\xi^2} = Z_3 \frac{N_0^2}{2\pi} \left( \frac{T}{\rho_s}\right)^2,
\label{magn}
\end{equation}
where $Z_3$ is a universal number and $N_0$ is the $T=0$ sublattice
magnetization
defined such that in the classical ($S\to\infty$) limit $N_0=S$, and
$S_0$ is defined such that $S_0=S(S+1)/3$ for $T\to\infty$.
The value of $Z_3$ can be estimated independently
using the numerical data for the two limiting cases $S=1/2$
\cite{EGSS,Greven:mc} and $S\to\infty$ \cite{Luscher:Weisz,THC}.

$Z_3$ is estimated by substituting
series expansion
data at the lowest temperature where calculation is possible,
$T_{\rm min}$, into Eq.(\ref{magn}). The data so-obtained for
$S=1/2$ and $S\to\infty$ differ by a factor of two:
$Z_3\approx 3.2$ for $S=1/2$ when estimated
at $T_{\rm min} = 0.35J$,
and  $Z_3\approx 6.6$ for $S\to\infty$ when estimated
at $T_{\rm min} \simeq  0.8JS^2$.
The disagreement of
the $S=1/2$ and $S\to\infty$ results for this universal parameter
suggests that at least one of the models
is not in the scaling limit at the respective $T_{\rm min}$ \cite{Z}.

Here we speculate on two possible causes of this discrepancy.
First, if the classical model is in the classical scaling limit
and the $S=1/2$ model is not in RC limit,
this discrepancy may be due to the vicinity of the classical to RC
crossover. In this case, $S_0/\xi^2T^2$ for the $S=1/2$ model should
increase at  temperatures lower than those studied numerically,
in order to reach its presumed larger scaling
limit for $T\to0$.
This scenario may be consistent with
neutron scattering measurements which give $S_0/\xi^2 \sim \mbox{const}$
\cite{Keimer:lco,Greven:scoc,Greven:S=1}.

Second, for either of the models
the true $T\to0$ scaling form for large $q\xi$ ($q$ is a deviation from
${\bf Q}=(\pi/a,\pi/a)$),
\begin{equation}
S({\bf q}) \simeq S_0 \frac{1+(B_f/2) \log(1+q^2\xi^2)}
{1+q^2\xi^2},
\label{scaling:q}
\end{equation}
and the hydrodynamic result, valid for $\xi^{-1}\ll q\ll T/c$,
\begin{equation}
S(q)\approx \frac{2}{3}\,
\frac{TN_0^2}{\rho_s q^2},
\label{hydro}
\end{equation}
cannot hold simultaneously at the temperatures studied
for any ${\bf q}$ inside the Brillouin
zone.
The reason is that the RC scaling form (\ref{scaling:q}), which has to match
the hydrodynamic expression
(\ref{hydro}) for $q\xi\to\infty$, is numerically found \cite{EGSS,THC}
to approach its asymptotic large-q limit
quite slowly. In the temperature range
studied, it still has large
deviations from this limit for any ${\bf q}$ inside the Brillouin
zone ($q\alt a^{-1}$), hence Eqs.(\ref{scaling:q}) and (\ref{hydro})
are not equal and at least one of them must be inconsistent with the data.

However, both Eqs.(\ref{scaling:q}) and (\ref{hydro})
are derived using essentially the same assumptions
that (i) $c/\xi\ll T$ and that (ii) $T$ is much smaller than other energy
scales. This
might indicate the importance of lattice corrections
(i.e.\ finite size of the Brillouin zone) to the ratio
$S_0/\xi^2$ in the temperature range studied,
in which $\xi$ is itself consistent with our proposed crossover
scaling scenario for any $S$.

In summary,
we present and analyze high-temperature series expansion data for
the spin-S square-lattice Heisenberg model for all $S$ in the range $S=1/2$
to $S=5/2$. In agreement with neutron scattering measurements,
we find that the correlation length deviates from the
low-temperature RC prediction of Eq.(\ref{HN}) for
$S>1/2$, and that the deviation becomes
larger for larger $S$. We suggest that this deviation primarily reflects a
crossover to classical  (as opposed to {\em renormalized}
classical, or RC) behavior in the temperature range $c/a\ll
T\ll\rho_s$, or equivalently $JS\ll T \ll JS^2$,
where the correlation length is nearly equal to that of
the classical $S\to\infty$ model.
In a subsequent publication \cite{SES}, which is now in preparation,
we propose to use comparisons between AFMs and FMs with the same value
of $S$ to study the role of quantum effects as a function of spin;
the results obtained in \cite{SES} using this method
are consistent with the crossover scenario proposed here.
Finally, we discuss the behavior
of the staggered structure factor $S_0$, which measures the strength
of magnetic correlations, and speculate on possible causes of its
apparent deviation from the $T\to0$ scaling limit.

We would like to thank S. Chakravarty, A.V. Chubukov,
B.I. Halperin, M.A. Kastner, M.S. Makivi\'{c}, D.R. Nelson, S.
Sachdev, and U.-J. Wiese for helpful discussions.
We thank J. Tobochnik and D.N. Lambeth for providing their
numerical data for comparisons.
A.S. is supported by an A.P.\ Sloan Research Fellowship.
R.R.P.S. is supported by the NSF under Grant No. DMR 93-18537.
The work at MIT was supported by the NSF under Grant No. DMR 93-15715 and by
the Center for Materials Science and Engineering under NSF Grant No. DMR
94-00334.


\end{document}